\documentclass[12pt]{article}

\def\rf#1{(\ref{eq:#1})}
\def\lab#1{\label{eq:#1}}

\def\br{\begin{eqnarray}}
\def\er{\end{eqnarray}}
\def\be{\begin{equation}}
\def\ee{\end{equation}}
\def\({\left(}
\def\){\right)}
\def\u2{\mid u\mid^2}
\def\rlx{\relax\leavevmode}
\def\IR{\rlx\hbox{\rm I\kern-.18em R}}

\usepackage{graphicx}

\begin{document}

\begin{titlepage}

\vspace*{-1cm}

\vskip 3cm

\vspace{.2in}
\begin{center}
{\large\bf Euclidean 4d exact solitons in a Skyrme type model}
\end{center}
\vspace{.3cm}

\begin{center}
L. A. Ferreira~

\vspace{.3 in}
\small

\par \vskip .2in \noindent
Instituto de F\'\i sica de S\~ao Carlos - IFSC/USP;\\
Universidade de S\~ao Paulo\\ 
Caixa Postal 369, CEP 13560-970, S\~ao Carlos-SP, Brazil\\

\vspace{.3cm}

and 


\par \vskip .2in \noindent
Instituto de F\'\i sica Te\'orica - IFT/UNESP\\
Universidade Estadula Paulista\\
Rua Pamplona 145\\
01405-900,  S\~ao Paulo-SP, Brazil\\

\normalsize
\end{center}

\vspace{.5in}

\begin{abstract}

We introduce a Skyrme type, four dimensional Euclidean field theory made of 
a triplet of scalar fields ${\vec n}$, taking values on the sphere
$S^2$, and an additional real scalar field $\phi$, which is dynamical
only on a three dimensional surface embedded in $\IR^4$. Using a
special ansatz we reduce the $4d$ non-linear
equations of motion into linear ordinary differential equations, which
lead to the 
 construction of an infinite number of exact soliton
solutions with vanishing Euclidean action. The theory possesses a mass 
scale which fixes the size of the solitons in way which differs from
Derrick's scaling arguments. The model may be relevant
to the study of the low energy limit of pure $SU(2)$ Yang-Mills
theory. 

\end{abstract}

\end{titlepage}

The aim of this Letter is to present a new four dimensional Euclidean
field theory and to construct an infinite number of exact soliton
solutions for it. The physical motivation of our work is
twofold. First, it contributes to the development of exact methods, in the
framework of integrable field theories, for the study of
non-perturbative aspects of physical theories \cite{afs97}. Second, 
it may connect to recent attempts to understand the role
of solitons in the low energy limit of pure Yang-Mills theories \cite{fndual}.

The Skyrme \cite{skyrme} and Skyrme-Faddeev \cite{sf} models are
effective field theories possessing topological solitons, with several
applications in various areas of physics. Such theories can not be
solved exactly, and those solutions are constructed through numerical
methods. The lack of an exact closed form for the solitons prevents
the understanding of several properties of the  phenomena
involved. It is therefore of physical interest to construct similar theories
possessing exact solutions.  In Ref. \cite{afz99} we have introduced a
model with the same field content and topology as the Skyrme-Faddeev
theory, and have constructed an infinite number of exact soliton
solutions with arbitrary topological Hopf charges. That model is
integrable in the sense of \cite{afs97} and possesses
an infinite number of local conservation laws. The exact solution was
obtained through an ansatz that, as explained in \cite{bf}, originates
from the conformal symmetry of the static equations of motion. 

In this Letter we propose a model with a Skyrme-Faddeev scalar field
taking values on the sphere $S^2$, together with an extra real scalar
field. An important aspect of the model is that, although defined on a
four dimensional Euclidean space $\IR^4$, the extra field lives only on
a three dimensional surface  embedded on $\IR^4$. That fact introduces
a mass parameter which fixes the size of the solitons, but in a way which
differs from Derrick's scaling arguments. The soliton solutions are
constructed through an ansatz, introduced in \cite{bf,afz99}, which
reduces the four dimensional non-linear equations of motion into
linear ordinary differential equations. The physical boundary
conditions are such that the Euclidean action vanishes when evaluated
on all soliton solutions.  
The theory can perhaps be
viewed as an effective theory for the pure $SU(2)$ Yang-Mills theory
in a way explained at the end of the Letter.  

The model considered in this Letter is defined by the action 
\be
S= M\, \int_{\Sigma} d\Sigma\; \frac{\partial_{\mu}\phi
  \partial^{\mu}\phi}{\(1+\u2\)^2} - 
\frac{1}{e^2} \int_{\IR^4}  d^4x \, H_{\mu\nu}^2
\lab{action}
\ee
where $\phi$ and $u$ are  real and complex scalar fields respectively,
with the latter related to the triplet of fields ${\vec n}$ living on $S^2$
(${\vec n}^2=1$) through the stereographic projection 
\be
{\vec n} = \(u+u^*,-i\(u-u^*\),\u2 -1\)/\(1+\u2\)
\lab{ndef}
\ee
$H_{\mu\nu}$ is the pull-back of the area form on $S^2$
\be
H_{\mu\nu}
=-2i\frac{\(\partial_{\mu} u\partial_{\nu} u^* - 
 \partial_{\nu} u \partial_{\mu} u^*\)}{\(1+\u2\)^2}  
=  {\vec n}\cdot\(\partial_{\mu}{\vec n} \wedge 
\partial_{\nu}{\vec n}\)
\lab{hdef} 
\ee
$M$ and $e^2$ are coupling constants, with $M$ having dimension of
mass and $e^2$ being dimensionless. $\Sigma$ is a $3$-dimensional
surface embedded in the four dimensional Euclidean space-time
$\IR^4$. As we explain below $\Sigma$ corresponds to the surface where
a special variable $y$ vanishes, and so one can write 
$\int_{\Sigma} d\Sigma = \int_{\IR^4}  d^4x \, \omega \,
\delta \(y\)$, with $\omega$ being an integration measure defined
below (see \rf{measure}).     

The action \rf{action} is not invariant under the $SO(3)$
transformations on the sphere $S^2$ defined by the fields ${\vec
  n}$. The first term of \rf{action} breaks that symmetry. However,
\rf{action} is invariant under the global $U(1)\otimes U(1)$ defined
by the transformations 
\be
u\rightarrow e^{i\alpha}\, u \qquad \qquad \phi
\rightarrow \phi + \beta
\lab{u1u1}
\ee

The Euler-Lagrange equations associated to \rf{action} are
\be
\partial^{\mu}\(H_{\mu\nu}\partial^{\nu}u\) +\frac{i}{4} \, Me^2\,
\omega \, \delta\(y\)\, \frac{u \(\partial \phi\)^2}{1+\u2}=0
\lab{equ}
\ee
together with its complex conjugate, and
\be
\partial^{\mu}\(\frac{\omega\delta\(y\)\partial_{\mu}\phi}{\(1+\u2\)^2}\)=0
\lab{eqphi}
\ee

Following \cite{afz99,bf} we introduce the ansatz
\be
u = \sqrt{\frac{1-g}{g}}\; e^{i\(m_1\varphi_1+m_2\varphi_2\)}\; ; \qquad 
\phi=\varepsilon_1\,m_1\varphi_1+\varepsilon_2\,m_2\varphi_2
\lab{ansatz}
\ee
based on a set of orthogonal curvilinear coordinates
$\(y,z,\varphi_1,\varphi_2\)$, with 
$-\infty<y<\infty$, $0\leq z \leq 1$, $0\leq \varphi_i\leq 2 \pi$,
$i=1,2$. We have that $g=g\(y,z\)$, $0\leq g \leq 1$,  with $m_i$,
$i=1,2$, being 
arbitrary integers, and $\varepsilon_i=\pm 1$, arbitrary signs.   

There are in fact two actions of
the type \rf{action} with different surfaces $\Sigma$, and as
consequence two sets of coordinates
$\(y,z,\varphi_1,\varphi_2\)$, admiting exact solutions through the ansatz
\rf{ansatz}. They are $4d$ generalizations of the $3d$ toroidal and $2d$
polar coordinates. The toroidal like coordinates are defined as
\br
x_1&=& \frac{r_0}{p}\, \sqrt{z}\, \cos \varphi_2 \qquad\qquad 
x_3= \frac{r_0}{p}\, \sqrt{1-z}\, \sin \varphi_1 \nonumber\\  
x_2&=& \frac{r_0}{p}\, \sqrt{z}\, \sin \varphi_2 \qquad \qquad \,
x_4= \frac{r_0}{p}\, \sinh y
\lab{torocoord}
\er
with $p=\cosh y - \sqrt{1-z} \cos \varphi_1$, and $r_0$ is a constant
with dimension of length.  The polar like coordinates are defined  as
\br
x_1&=& r_0\, e^y \, \sqrt{z}\, \cos \varphi_2 \qquad\qquad 
x_3= r_0\,e^y\, \sqrt{1-z}\, \cos \varphi_1 \nonumber\\  
x_2&=& r_0\, e^y\, \sqrt{z}\, \sin \varphi_2 \qquad \qquad\,
x_4= r_0 \, e^y\, \sqrt{1-z}\, \sin \varphi_1
\lab{polarcoord}
\er  
The metric is given by  
\be
ds^2= \frac{r_0^2}{\tau^2}\(  dy^2+
\frac{dz^2}{4\, z\(1-z\)}+\(1-z\) d{\varphi_1}^2+ 
z\, d{\varphi_2}^2 \)
\lab{metric}
\ee
with $\tau=p$ in the toroidal case, and $\tau=e^{-y}$ in the polar case. 
The surface $\Sigma$ is defined by the equation $y=0$, and in the
toroidal case is 
\be
\Sigma_T \equiv \{ x^{\mu} \in \IR^4 \mid x_4=0 \quad {\rm or} \quad
r^2\rightarrow \infty\}
\ee
with $r^2=x_1^2+x_2^2+x_3^2+x_4^2$. In the polar case we have
\be
\Sigma_P \equiv S^3 = \{ x^{\mu} \in \IR^4 \mid r^2 = r_0^2\}
\ee
The integration measure in the toroidal and polar cases are
respectively
\be
\omega_T = p^2/r_0\; , \qquad \omega_P = e^{-2y}/r_0
\lab{measure}
\ee

Due to the orthogonal character of the coordinates and the form of the
ansatz \rf{ansatz} it turns out that the gradient of $\phi$ is
orthogonal to the gradients of $\mid u\mid$ and $y$. In the toroidal
case one can easily check that
$\omega\partial^2\phi=-\partial_{\mu}\omega\partial^{\mu}\phi=-2
\varepsilon_1\,m_1 p^3 
\sin \varphi_1/\(r_0^3\sqrt{1-z}\)$. In the polar case one has 
$\omega\partial^2\phi=-\partial_{\mu}\omega\partial^{\mu}\phi=0$. Therefore,
in both cases the equation of motion \rf{eqphi} for $\phi$  is automatically
satisfied by the ansatz \rf{ansatz}.  

Replacing the ansatz \rf{ansatz} into the equation of motion \rf{equ} 
 one gets the same equation for the profile function
$g\(y,z\)$ for the two types of coordinates \rf{torocoord} and
\rf{polarcoord}. It is a {\em linear} partial differential equation given by  
\be
4\,\frac{z\(1-z\)}{\Omega}\,\partial_z\(\Omega \,\partial_z g\) + 
\partial_y^2 g + \frac{Me^2r_0}{8}\,\delta\(y\)\, g=0
\lab{eqg}
\ee
with $\Omega=m_1^2\, z + m_2^2 \, \(1-z\)$.
It can be solved by separation of variables
\be
g\(y,z\) = k\(y\)\, h\(z\)
\ee
giving the linear ordinary differential equations
\be
k^{\prime\prime}\(y\) - \lambda^2\, k\(y\) +
 \frac{Me^2r_0}{8}\,\delta\(y\)\, k\(y\) = 0
\lab{eqk}
\ee
and 
\be
z\(1-z\)\partial_z\( \( q^2 z+\(1-z\)\)\partial_z h\(z\)\) + 
\frac{\lambda^2}{4}\, \(  q^2 z+\(1-z\)\) h\(z\) = 0
\lab{eqh}
\ee
where $\lambda^2$ is the separation of variables constant, and 
\be
q=\frac{\mid m_1\mid}{\mid m_2\mid}
\lab{qdef}
\ee
We must have $0\leq g\leq 1$, and so the solutions of \rf{eqk} and
 \rf{eqh} should not diverge. Since any solution of those equations can be
 rescaled by a constant, we can use  any finite solution  with
 definite sign, to
 construct a $g$ satisfying that requirement. 

Using the integral representation for the Dirac delta function one gets that
 the solution for \rf{eqk} is  
\be
k\(y\)= k\(0\) \, \frac{Me^2r_0}{16 \pi}\, \int_{-\infty}^{\infty} d\nu 
\frac{e^{i\nu\,y}}{\nu^2+\lambda^2}
\lab{preksol}
\ee
From Cauchy's theorem one gets 
\be
\int_{-\infty}^{\infty} d\nu 
\frac{e^{i\nu\,y}}{\nu^2+\lambda^2} = \frac{\pi}{\mid \lambda\mid} 
e^{-\mid \lambda\mid\mid y\mid}
\lab{cauchy}
\ee
Therefore, the consistency of \rf{preksol} and \rf{cauchy} at $y=0$
 fixes the value of $\lambda$ to
\be
\mid \lambda\mid= \frac{Me^2r_0}{16}
\lab{fixlambda}
\ee
and the solution for $k$ takes the form
\be
k\(y\)= k\(0\)\, e^{-Me^2r_0\mid y\mid/16}
\lab{ksol}
\ee
Notice that both integrands in \rf{action} are positive definite. The
 sign of the measure of the $\Sigma$ integration is dictated,
 according to \rf{measure},  by the sign of $r_0$. Therefore, the
 relative sign of the two terms in \rf{action} is given by the sign of
 $(-Me^2r_0)$, which due to \rf{fixlambda} must be negative. So, the
 Euclidean action \rf{action} can not be chosen to be positive definite. 

The equation for $h$ \rf{eqh} is what is called a Heun equation
\cite{heun}. Having four regular singular points, it is a
generalization of the Gauss' hypergeometric equation. Indeed, for $q=1$
eq. \rf{eqh} reduces to the  hypergeometric equation. 
In our case, the
fourth singular point will not be a concern  since $\( q^2
z+\(1-z\)\)$ can only vanish at $z=1$
for $q=0$, and we will not treat that case.  

The boundary conditions that $h\(z\)$ has to satisfy will be dictated
by the requirement of finite action solutions. Replacing the ansatz
\rf{ansatz} into the action \rf{action} one gets for both coordinates
\rf{torocoord} and \rf{polarcoord} that 
\be
S= -\frac{16\pi^2}{e^2}\int_{-\infty}^{\infty}dy  \int_{0}^1 dz 
\frac{\Omega}{z\(1-z\)} \(4 z\(1-z\)\(\partial_z g\)^2 
+  \(\partial_y g\)^2 -\frac{Me^2r_0}{8}\, \delta\(y\) \, g^2\)
\ee
Using \rf{eqg} multiplied by $g$, integrating by parts and using the
solution \rf{ksol}, one gets
\be
S=-\frac{\(32\pi k\(0\)\)^2}{Me^4r_0}\(m_1^2\, h\, h^{\prime}\mid_{z=1}-
m_2^2\, h\, h^{\prime}\mid_{z=0}\)
\lab{fullactionborder}
\ee
If a solution of \rf{eqh} does not vanish at $z=0$ or $z=1$ then its
first derivative has a logarithmic divergence there. So, in order to
have finite action one needs $h$ to vanish  at $z=0$ and $z=1$. In
such case, the action vanishes. In addition, since we need $0\leq
g\leq 1$, and since the solution for $k$ given in \rf{ksol} has a
definite sign, it follows that $h$ can not change sign between $z=0$
and $z=1$. Therefore, the boundary conditions required for $h$ are
\be
h\(0\)=h\(1\)=0\; ; \qquad \mbox{\rm $h\(z\)\neq 0$ for $0<z<1$}
\lab{boundcond}
\ee
Notice that \rf{eqh} is invariant under the joint transformations
$z\leftrightarrow \(1-z\)$ and $q \leftrightarrow 1/q$. Therefore, if 
$h_q\(z\)$ is a solution of \rf{eqh} for a
given value of $q$, then 
\be
h_{1/q}\(z\)\equiv h_q\(1-z\)
\lab{invqrel}
\ee
is also a solution of \rf{eqh} for the inverse value of $q$. We will then
construct solutions for \rf{eqh} in power series around $z=0$ for
$0<q\leq 1$. The solutions for $q>1$ are obtained from those through
\rf{invqrel}. Replacing $h\(z\) = \sum_{n=0}^{\infty} c_n z^n$ into
\rf{eqh} we get that  
$c_0=0$ and  $c_1$ is arbitrary. The remaining coefficients are
obtained through the recursion relation
\br
c_n &=& \frac{-1}{n\(n-1\)}\left[ 
\(q^2-1\)\(\frac{\lambda^2}{4} -
\(n-2\)^2\)\, c_{n-2}
\right. \lab{recursion}\\
&+& \left. 
\(\frac{\lambda^2}{4}-\(n-1\)\(n-2\) 
+ \(q^2-1\) \(n-1\)^2\)\, c_{n-1}
\right] \nonumber
\er
Therefore, $h\(z\)$ vanishes at $z=0$, as required by
\rf{boundcond}. For a given value of $q$ the locations of the other
zeroes of $h$ depend upon the value of $\lambda$. We then have to tune
$\lambda$ in order for the next zero to fall exactly at $z=1$. The
remaining zeroes will then happen for $z>1$, as required by
\rf{boundcond}. That gives a discrete
spectrum for $\lambda$ and the size of the solution is fixed by the
length scale $r_0$ through \rf{fixlambda}. For $q=1$ the solution for
such a problem is exact, since the series solution truncates, and we get 
\be
h_{q=1}\(z\)= 4 \, z\,\(1-z\) \qquad \qquad \qquad \lambda_{q=1}^2=8
\ee
For other values of $q$ the series does not truncate and we can tune
$\lambda$ numerically. The values of $\lambda$, for some values of
$q$, such that $h\(1\)=0$ is given in Table
\ref{lambdatable}. Due to \rf{invqrel} the values of $\lambda$
satisfying the boundary conditions \rf{boundcond} are the same for $q$
and $1/q$.

\begin{figure}[hbt]
\scalebox{1.1}{
\includegraphics{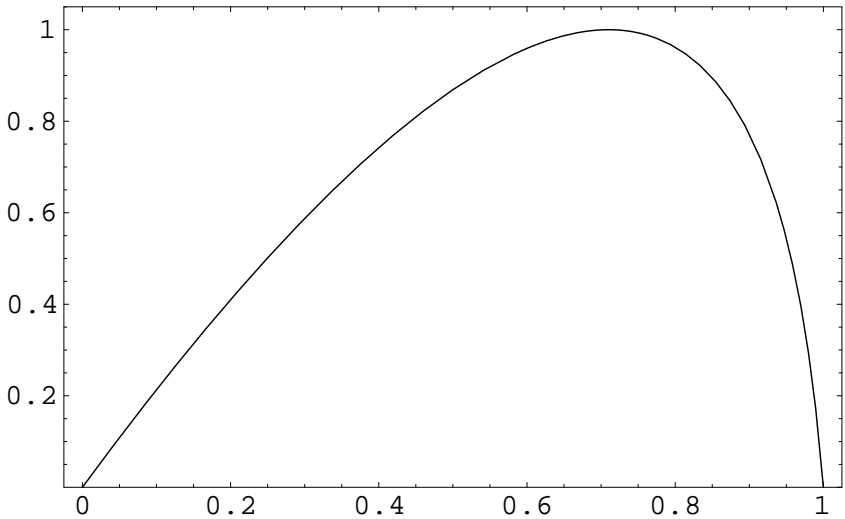}}
\caption{\label{fig:h16} Plot of
	$h_{q=1/6}\(z\)$, normalized as in \rf{hqnorm}. The form of 
	$h_q$ for other values of $q$  is similar to this one. The
	peak moves to the right as $q$ decreases. The locations $z_c$
	of the peaks are given in Table \ref{lambdatable}.  The plots of
	$h_q$ and $h_{1/q}$ are 
	related by reflection around $z=1/2$ (see \rf{invqrel}).}
\end{figure}

As we commented below \rf{ansatz} the profile function $g$ has to take
values between zero and unity. Therefore, we take $k\(0\)=1$
in \rf{ksol}. In addition, we  normalize $h\(z\)$ such that its
maximum value, on the 
interval $0\leq z \leq 1$, is unity. We then write 
\be
h_q\(z\)= c\(q\)\, \(z + \sum_{n=2}^{\infty} c_n z^n\)
\lab{hqnorm}
\ee
with $c\(q\)$ given in Table \ref{lambdatable}. The $c_n$'s are determined
from \rf{recursion} with $c_0=0$ and $c_1=1$. The position $z_c$ of the
maximum, i.e. the zero of  the derivative ($h_q^{\prime}\(z_c\)=0$), is
also given in Table \ref{lambdatable}. The form of the functions
\rf{hqnorm} is exemplified in the Figure \ref{fig:h16}
for the case $q=1/6$.  Consequently, the solutions
for the profile function $g$ become    
\be
g_{q=1}\(y,z\) = 4 \, z\,\(1-z\)\, e^{-2\sqrt{2}\,\mid y\mid}
\qquad \qquad \qquad 
g_q \(y,z\)= h_q\(z\)\, e^{-\mid \lambda\(q\)\mid\mid y\mid}
\lab{finalsolg}
\ee
with $\lambda$ fixed by \rf{fixlambda} and the boundary conditions
\rf{boundcond}. Values of $\lambda$ for some $q$'s are given in Table
\ref{lambdatable}.

\begin{table}[thb]
\begin{tabular}{|c||l|r|l|l|}
\hline
$q$ & $\qquad\;\;\;\lambda^2/4 $ & $\Lambda\(q\)\quad\; $ &  $\qquad z_c$ &
$\qquad c\(q\)$\\ 
\hline \hline
1 &    2  &                      7.542472   &      0.5  &               4\\
\hline
9/10 & 1.99778564857563(7) &     7.568858   &   0.5131694374 & 3.8013381236\\
\hline
4/5 &  1.99015726516381(9)&      7.661008   &   0.5278864055 & 3.6045699997\\
\hline
3/4 &  1.98376526603091(4)&      7.739752   &   0.5359452134 & 3.5064351049\\
\hline
2/3 &  1.96834721693012(0)&      7.935672   &   0.5506341233 & 3.3423893780\\
\hline
1/2 &  1.91364355092039(7)&      8.707347   &   0.5862821184 & 3.0065092866\\
\hline
2/5 &  1.85959184446300(8)&      9.612200   &   0.6134572404 & 2.7934885325\\
\hline
1/3 &  1.81146288327809(9)&     10.571742   &   0.6351236762 & 2.6427573006\\
\hline
1/4 &  1.73364482996707(3)&     12.547803   &   0.6678866103 & 2.4390777756\\
\hline
1/5 &  1.67496824177613(6)&     14.538513   &   0.6917764653 & 2.3049609403\\
\hline
1/6 &  1.62948847681927(1)&     16.519869   &   0.7101532977 & 2.2084486426\\
\hline
1/10 & 1.51764973827612(1)&     24.281837   &   0.7557187711 & 1.9890631347\\
\hline
1/100& 1.25301610750(6)&       177.9(3) $\quad$ 
                                            &   0.8719052146 & 1.5135800082\\ 
\hline
1/1000&1.1610118(7) &                       &   0.9168161751 & 1.3465808941\\
\hline
\end{tabular}
\caption{Numerical results for the solutions:   
   {\em i)} Some chosen values of $q$ (see definition in \rf{qdef}) are
   given on the first column;  {\em ii)} On the second column are the
   values of $\lambda$ such that $h_q$ vanishes  at
   $z=1$ (see below \rf{recursion}); {\em iii)} $\Lambda\(q\)$, given
   on the third column, is defined in \rf{lambdaqdef}; 
   {\em iv)}  $z_c$ given on the
   fourth column is such that $h_q^{\prime}(z_c)=0$ and $h_q(z_c)=1$;
   {\em v)} $c(q)$ given on the fifth column is the normalization of
   $h_q$ (see \rf{hqnorm}) such that its maximimum value is unity on
   the interval $0\leq z\leq 1$. The numbers are exact up the digit
   shown, except for those cases where the first uncertain digit is
   shown between parenthesis.}
\label{lambdatable}
\end{table}

Since the solutions \rf{finalsolg} satisfy the boundary conditions
\rf{boundcond},  
the action \rf{action}, evaluated on them, vanishes (see
\rf{fullactionborder}). However, both 
terms of the action are finite and cancel each other. We can then
introduce an {\em energy} function $E$, which is the same for the two
coordinates systems \rf{torocoord} and \rf{polarcoord}, and it is
given by  
\be
E\equiv M\, \int_{\Sigma} d\Sigma\; \frac{\partial_{\mu}\phi
  \partial^{\mu}\phi}{\(1+\u2\)^2} = 
\frac{1}{e^2} \int_{\IR^4}  d^4x \, H_{\mu\nu}^2 
= \frac{32\, \pi^2}{e^2}  \mid m_1 \, m_2 \mid  \Lambda \(q\)
\ee
with
\be
\Lambda \(q\) = \Lambda \(\frac{1}{q}\) = \mid \lambda\mid \,
\int_0^1 dz \, \( \frac{q}{1-z}+\frac{1}{q\, z}\) \,h_q^2\(z\)
\lab{lambdaqdef}
\ee
The values of $\Lambda \(q\)$ for some of values of $q$ are given in
Table \ref{lambdatable}. 
Notice that $E$ does not depend upon the signs of the integers $m_i$,
and it is invariant under the interchange $m_1\leftrightarrow m_2$.  

The solitons we have obtained in this Letter are smooth solutions with
vanishing Euclidean 
action, localized around the three dimensional surface $\Sigma$. That
surface plays the role of a source for the solution. The eq. \rf{equ} with
the delta term, indicates that character. In addition, since the $O(3)$
symmetry is broken to $U(1)\otimes U(1)$, the theory \rf{action} can perhaps be
seen as an effective theory.  
There can be many applications in that direction, but perhaps one that
is worth pursuing is the following. 
Consider pure $SU(2)$ Yang-Mills theory (without matter) and the Faddeev-Niemi
\cite{fndual} parametrization of the physical degrees of 
freedom of the gauge field (a modification of Cho's ansatz
\cite{cho})  
\be
{\vec A}_{\mu} = C_{\mu} \, {\vec n} + 
\partial_{\mu}{\vec n}\wedge {\vec  n} + \rho\, \partial_{\mu} {\vec n} + 
\sigma \,  \partial_{\mu}{\vec n}\wedge {\vec  n}
\ee
where $\rho$ and $\sigma$ are real scalar fields, $C_{\mu}$ is an
abelian gauge field, and ${\vec  n}$ is the same as in \rf{ndef}. The
arrows stand for the orientation of the fields in the algebra of
$SU(2)$. One can use
the scalar fields to construct a complex field $\Phi \equiv \rho + i
\sigma \equiv \mid \Phi \mid e^{i \phi}$. It could happen that as the
energy decreases the degrees of freedom associated to $C_{\mu}$ and
$\mid \Phi \mid$ get frozen. The 
 phase $\phi$ of $\Phi$ however gets frozen at different values in different
 regions of space. On the border of such regions it could still have
 some dynamics. We would then have at low energies, the degrees of
 freedom of the Yang-Mills field being described by the field ${\vec  n}$,
 all over the space, plus the phase $\phi$ on the border of those
 regions. Very probably the effective action describing such regime is
 very complex, but the action \rf{action}  could perhaps describe, in
 Euclidean   space, some aspects of such phase of the Yang-Mills theory.

\vspace{2cm}

{\bf Acknowledgements:} I am very indebted to Olivier Babelon for
numerous helpful and iluminating discussions throughout all phases of this
work. I am also grateful to helpful discussions with F. C. Alcaraz,
H. Aratyn, C. Biasi, J.F. Gomes,  M. Speraficco, and A.H. Zimerman. I
thanks the 
hospitality at LPTHE-Paris where part of this work was developed. The
work has been funded  by the Capes/Cofecub collaboration,  Fapesp (Projeto
Tematico) and a CNPq research grant.  

\vspace{1cm}

\end{document}